\documentclass[twocolumn,iop,preprint]{emulateapj}%
\usepackage{graphicx}
\usepackage{amsmath}
\usepackage{amsfonts}
\usepackage{amssymb}%
\providecommand{\U}[1]{\protect\rule{.1in}{.1in}}

\usepackage{color}

\begin{document}

\title{The anatomy of a long gamma-ray burst: a simple classification scheme for the emission mechanism(s).}
\shorttitle{Classification of long GRBs}
\author{D. B\'egu\'e \altaffilmark{1,2,$\ddag$} and J. Michael Burgess \altaffilmark{1,2,$\dag$}}
\shortauthors{B\'egu\'e \& Burgess}
%\altaffilmark{1,2,3} et al.

\altaffiltext{1}{The Oskar Klein Centre for Cosmoparticle Physics,
  AlbaNova, SE-106 91 Stockholm, Sweden}
\altaffiltext{2}{Department of Physics, KTH Royal Institute of Technology, AlbaNova University Center, SE-106 91 Stockholm, Sweden}
\altaffiltext{$\dag$}{jamesb@kth.se}
\altaffiltext{$\ddag$}{damienb@kth.se}

\begin{abstract}

  Ultra-relativistic motion and efficient conversion of kinetic energy
  to radiation are required by gamma-ray burst (GRB) observations, yet
  they are difficult to simultaneously achieve. Three leading
  mechanisms have been proposed to explain the observed emission
  emanating from GRB outflows: radiation from either relativistic
  internal or external shocks, or thermal emission from a photosphere.
  Previous works were dedicated to independently treating these three
  mechanisms and arguing for a sole, unique origin of the prompt
  emission of gamma-ray bursts. In contrast, herein, we first explain why 
  all three models are valid mechanisms and that a contribution
  from each of them is expected in the prompt phase. Additionally, we
  show that a single parameter, the dimensionless entropy of the GRB
  outflow, determines which mechanism contributes the most to the
  emission. More specifically, internal shocks dominate for low values
  of the dimensionless entropy, external shocks for intermediate
  values and finally, photospheric emission for large values. We
  present a unified framework for the emission
  mechanisms of GRBs with easily testable predictions for
  each process.

\end{abstract}
\maketitle

\section{Introduction}

Gamma-ray bursts (GRBs) are the most luminous sources in the Universe,
and yet, their emission mechanism(s) is still lacking a unique
explanation, for reviews see  \cite{M06, P15, KZ15}. Observations of rapid temporal variability and extremely
high-energy photons require GRBs to have ultra-relativistic motion of
their outflow plasma with Lorentz factors at least as high as a few
hundred \citep{PS93}, implying that most of the energy of the burst is
in kinetic form. Such large Lorentz factor are also indirectly suggested
by the non-observation of neutrinos in coincidence with GRBs \citep{AAA12}
and the constraints from the early optical afterglow, see \textit{e.g.}
\cite{RAA09}.

Therefore, efficient conversion of kinetic energy to
radiation is required for any model attempting to explain the extreme
isotropic\footnote{The GRB emission is expected to be collimated in a narrow jet. However,
in the following, we only deal with isotropic-equivalent quantities.}
luminosities ($\gtrsim10^{51}$ erg s$^{-1}$) produced in GRBs. The
leading mechanisms are the internal shock model \citep{RM94,DM98}
(hereafter IS), the external shock model \citep{RM93,CD99} (hereafter
ES), and photospheric models \citep{G86,P86} (hereafter PE). The first
one assumes an \textit{unsteady} outflow. Rapid variations in
the outflow's Lorentz factor result in internal collisions, converting
kinetic energy to synchrotron radiation of accelerated electrons. On
the other hand, the ES  scenario considers that the relativistic outflow is
decelerated by an external circum-burst medium (hereafter CBM). Finally, PE
models rely on efficient emission by the plasma, as it becomes
optically thin when the plasma density and temperature drop due to the outflow's
expansion. 

Several improvements were considered for each of these models in order
to allow them to explain all bursts. However, it appears that they all
have problems to explain certain specific features in specific
bursts. Among them, GRB~090902B, with its nearly thermal spectrum,
cannot be reconciled with either IS nor ES \citep{RAZ10}, and the very smooth
pulsed GRB~141025A which %can not be reconciled with neither IS nor PE, but 
is well-explained by ES \citep{BBR15}.

Still, all three models have the potential to explain several features
of GRBs. In fact, all three emission mechanisms are expected in \textit{any}
GRB scenario. Only their relative luminosities have
to be quantified and the delay (for an observer) between them
constrained. \cite{S97} considered the situation of an early onset of
the afterglow (due to an external shock), which leads to mixed
emission from an unspecified mechanism during the prompt phase of the
GRB and the external shock. More recently, this idea was also put
forth by \cite{KB09}, who tried to explain the late onset of the GeV
emission with synchrotron emission of electrons accelerated at the
external shock.

%{\color{red} In this paper, we consider the expansion of a thermal fireball,  and}
%we propose to characterize the emission
%mechanisms solely based upon the dimensionless entropy of the outflow
%$\eta \equiv E_{\rm tot}/(M c^2) \gg 1$, where $E_{\rm tot}$ is the total
%energy of the burst, $M$ is the total baryonic mass of the outflow,
%and $c$ the speed of light.

 In this paper, we consider the expansion of a classical thermal fireball \citep{G86,P86,PSN93},
for which magnetic fields are sub-dominants. There also exist magnetic acceleration models based on the spreading of magnetic lines,
see \textit{i.e.} \cite{NMF07} or magnetic reconnection \citep{DS02,G06}. However, these models might be challenged by
observations, see \cite{BGP15,BP15}.  
In addition, the classical thermal fireball model neglects the interaction between the progenitor of the GRBs,
thought to be a massive star undergoing collapse \citep{W93,WB06}. However, numerical simulations have shown that, if
the central engine remains active long enough, the expansion of the jet is mostly unperturbed  once
it has open a funnel in the progenitor \citep{LMB12}. Oblique shocks can also stall the initial expansion of the
jets \citep{TMR07,IRA13}, but here we neglect this effect.

 In this paper, we propose to characterize the emission
mechanisms solely based upon the dimensionless entropy of the outflow
$\eta \equiv E_{\rm tot}/(M c^2) \gg 1$, where $E_{\rm tot}$ is the total
energy of the burst, $M$ is the total baryonic mass of the outflow,
and $c$ the speed of light.

First, we review all three models  in their unembellished version and their main
characteristics. Second,  we come to the point of the paper and we show how the value of the dimensionless
entropy strongly constrains all three emission mechanisms. Then, we analyse the
implications and predictions of our classification on the afterglow
and other GRB properties. Finally, we demonstrate how our
classification can be tested against observations.

\section{Emission models}
The ``classical" fireball model \citep{P90,PSN93} assumes that a large
quantity of energy $E_{\rm tot} \sim 10^{53} \text{erg}$ is released
by an unspecified cataclysmic event such as the death of a hyper-massive
star. From the millisecond variability observed in a few GRBs
(but not all, see \cite{GB14}), the origin of the outflow is
constrained to within a few $10^{8}$ cm from the center of the
progenitor. Such a large amount of energy in such a small region leads
to the creation of an optically thick plasma. Due to its high thermal
pressure, the outflow expands and is accelerated to relativistic
speeds. The expansion can be described by two phases
\citep{PSN93}. During the initial accelerating phase, the Lorentz
factor of the outflow increases proportionally to the radius, while
after (eventually) reaching its limiting Lorentz factor $\Gamma
 \lesssim \eta$, the outflow coasts at constant velocity.

Other acceleration models based on magnetic fields are currently
strongly debated in the literature \citep{DS02,NMF07}. However, our
results can be re-parametrized to the magnetization of the outflow,
which plays a comparable role to $\eta$ for the expansion
dynamics. Therefore, the results are expected to be \textit{qualitatively}
 similar.

\subsection{Photospheric model}

In a photospheric model, the thermal energy is released when the
outflow becomes transparent at typical radii $R_{\rm ph} \sim
10^{10}-10^{12}$ cm,  for a recent review on photospheric emission, see \cite{V14}.
The efficiency of the photospheric emission, assuming no dissipation, can
be evaluated as the ratio of the thermal energy emitted at the
photosphere $E_{\rm th}$ to the total energy $E_{\rm tot}$

\begin{align}
  \epsilon_{\rm ph}  &= \frac{E_{\rm th}}{E_{\rm tot}}  \sim 1- \frac{\Gamma_{\rm ph}}{\eta}, 
\end{align}
%%\frac{E_{\rm tot}-E_{\rm kin}}{E _{\rm tot}}  \sim 1- \frac{\Gamma_{\rm ph} M c^2}{\eta M c^2} \nonumber \\

\noindent where  $\Gamma_{\rm
  ph}$ is the Lorentz factor at the photosphere. Therefore, a bright
photosphere requires  $\Gamma_{\rm ph} \ll \eta$, implying
transparency of the outflow in the initial accelerating phase or in
the transition phase between accelerating and coasting phases. The
limiting $\eta$ value separating photospheric emission in the
accelerating phase from transparency in the coasting phase is  \citep{RM94, T94} 

\begin{align}
  \eta^{*} & = \left (\frac{\sigma_T E_{\rm tot}}{4 \pi m_p c^3 \Delta t R} \right )^{\frac{1}{4}} \nonumber \\
                 & \sim 7 \times 10^{2} \text{~~} E_{53}^{\frac{1}{4}} R_{8}^{-\frac{1}{4}} \Delta t_{\rm 5s}^{-\frac{1}{4}},
\end{align}

\noindent where $R$ is the radius at which the outflow starts to
accelerate, $\Delta t$ is the time the central engine remains active,
$\sigma_{\rm T}$ is the Thompson cross-section, $m_{\rm p}$
is the proton mass. For all parameters but $\Delta t$, which is normalized
to $5\text{s}$ (see below), we adopt the
convention $X=10^n X_n$, where all quantities $X$ are in CGS units. When
$\eta > \eta^{*}$, more than half of the energy of the burst is
emitted at the photosphere. Therefore, in this case, we consider that
the emission from the photosphere dominates the emission of the prompt
phase, regardless of which mechanism is responsible for any remaining
emission. We further restrain the study to $\eta < \eta^{*}$ and consider
the emission be dominated by photospheric emission for $\eta > \eta^{*}$.

%DISSIPATION ?

In first approximation, as soon as the outflow becomes transparent, the radiative
pressure decreases abruptly, stopping the acceleration of the
outflow.\footnote{ Here, we assume that the Compton drag is negligible, as implied by $\eta < \eta^{*}$, see equations (10) and (11) of \cite{MR00}. However, if $\eta \gg \eta^{*}$, then the flow is still accelerated above the photosphere. See a complete discussion in \cite{GW98} and \cite{MR00}. In the following we do not consider this effect, as we consider $\eta < \eta^{*}$.} This implies that the kinetic energy of the blast wave above
the photosphere is set by $\Gamma_{\rm ph}$.
% Obviously, when the outflow is in the coasting phase, $\Gamma_{\rm ph} \lesssim \eta$ and $E_{\rm kin} \lesssim E_{\rm tot}$. 
%Also faster outflows becomes transparent earlier (at constant energy), leading to a more intense photospheric emission.
Neglecting high-latitude effects, the duration of the photospheric emission is roughly given by the
light crossing time of the outflow $\Delta t_{\rm ph} \sim \Delta t \sim l/c \sim
\text{few seconds}$, independent of the value of the Lorentz
factor\footnote{This is true if high-latitude effects are neglected. They might
be identified as the flux and temperature of the blackbody decreases as
$F_{\rm BB} \propto t^{-2}$ and $T_{BB} \propto t^{-2/3} $, see \cite{P08}.}. Here $l \sim c \Delta t$ is the laboratory width of the
outflow. %, which is of main importance for the following
         %discussion. Here $l \sim c \Delta t$ is the laboratory width
         %of the outflow. As the expansion continues, internal shocks
         %become efficient.
The luminosity at the photosphere is then  approximated by:

\begin{align}
L_{\rm ph} & = \epsilon_{\rm ph} \frac{E_{\rm tot}}{\Delta t}.
\end{align}

 An important assumption in this derivation, is that, if dissipation occurs below the photosphere, it only amounts for
a few percent of the total energy $E_{\rm tot}$. This is in agreement with the analysis of spectra in the guise of a photospheric model,
see \textit{i.e.} \cite{ALN15}. In addition, it was demonstrated in \cite{BI14} that if dissipation amounts for a large fraction of the total energy,
the resulting observed photospheric peak energy would be too low (around 1keV) to corresponds to the peak energy of GRBs or to the additional black-body
component identified in some bursts \citep{GCG03,R04}.
%\noindent As the expansion continues, IS become efficient.

\subsection{Internal shock model}

As the expanding outflow is unlikely to be steady due to small
variations in the wind parameters, sections with different speeds will necessarily
collide with one another. These collisions (internal shocks) convert a
fraction of the kinetic energy to internal energy, which can
subsequently be radiated away by accelerated electrons. The collided sections will
then form a single merged system. These collisions take place at
typical radii\footnote{One can imagine collisions at smaller radii even below the photosphere, see \citep{RM05}. However, here we consider pure internal shocks as in their original definition.} $R_{\rm IS} \sim 10^{14}$ cm. In this section, we follow the
derivation of \cite{DM98} to obtain qualitative
estimates. Assuming a steady injection mass rate and a Lorentz factor
variation $\Delta \Gamma$, the Lorentz factor of the merged system
after an internal shock $\Gamma_{\rm IS} $ is  \citep{DM98} 

\begin{align}
\Gamma_{\rm IS} = \sqrt{\Gamma_{\rm ph} (\Gamma_{\rm ph} - \Delta \Gamma) } \sim (1-\frac{\Delta \Gamma}{2 \Gamma_{\rm ph}}-\frac{\Delta \Gamma^2}{8 \Gamma_{\rm ph}^2}) \Gamma_{\rm ph}\text{,} 
\end{align}

\noindent where the last equality is obtained for $\Delta \Gamma  < 
\Gamma_{\rm ph}$.

The efficiency of this process can be estimated as \citep{KPS97} 

\begin{align}
\epsilon_{\rm IS} & = \frac{E_{\rm kin} - E_{\rm kin, IS}}{E_{\rm kin}} &= 1 - \frac{ 2\Gamma_{\rm ph} \left ( 1- \frac{ \Delta \Gamma}{2 \Gamma_{\rm ph}}-\frac{\Delta \Gamma^2}{8 \Gamma_{\rm ph}^2} \right ) }{ (2 \Gamma_{\rm ph}-\Delta \Gamma)}  \nonumber \\
& \sim \frac{\Delta \Gamma^2}{8 \Gamma_{\rm ph}^2}
\label{eq:epsilonIS}
\end{align}

\noindent Here $E_{\rm kin, IS} = (1-\epsilon_{\rm ph}-\epsilon_{\rm
  IS})E_{\rm tot}=\epsilon E_{\rm tot}$ is the kinetic energy of the outflow after
dissipation of energy by IS. This hydrodynamical efficiency should be
multiplied by an additional factor $\xi_{\rm rad, IS} < 1$ to obtain
the radiative efficiency. Then the luminosity of an IS is roughly given
by:%This estimation is only the hydrodynamical efficiency and not the
   %radiative efficiency, which is obtained by multiplying
   %Eq.(\ref{eq:epsilonIS}) by an additional factor $\xi_{\rm rad, IS}
   %< 1$. The luminosity is estimated by:

\begin{align}
L_{\rm IS} = \xi_{\rm rad, IS} \frac{\epsilon_{\rm IS} (1-\epsilon_{\rm ph})E_{\rm tot}}{\alpha_{\rm IS} \Delta t}, \label{eq:lumIS}
\end{align}

\noindent where $\alpha_{\rm IS} > 1$ is a numerical factor of order
unity, which takes into account that the IS duration for the observer is slightly
larger than the light crossing time of the outflow, see \cite{DM98}.

For efficient internal shocks, i.e., $\Delta \Gamma \simeq \Gamma_{\rm ph}$,
with high $\eta$, it is clear that the outflow must be very unsteady.
%i.e., with large $\Delta \Gamma$. %When $\eta$ increases, efficient internal shock requires a very
%unsteady outflow with $\Delta \Gamma \sim \Gamma$. 
Moreover, if the acceleration is incomplete at the photosphere
($\Gamma_{\rm ph} \ll \eta$), the efficiency does not increase since
$\Delta \Gamma$ and the Lorentz factor scale proportionally to the
radius in the accelerating phase.

The strength of the model is its ability to reproduce the observed
highly variable light curves of GRBs, down to the millisecond
time-scale. However, the model has a low radiative efficiency on the
order of a few percent \citep{KPS97}, implying that the emission can
be out-shined by the photospheric emission \citep{DM02}, but also by
an external shock, which will necessarily result as the outflow
expands and interacts with the CBM.
% A lower limit
%on the duration on which photon from IS are observed is given by the light-crossing time of the outflow \cite{DM98}. Further expansion of the outflow leads to its efficient interaction with the CBM via external shock.

\subsection{External shock model}

\begin{figure}[t]
\centering
\includegraphics[width=0.4\textwidth]{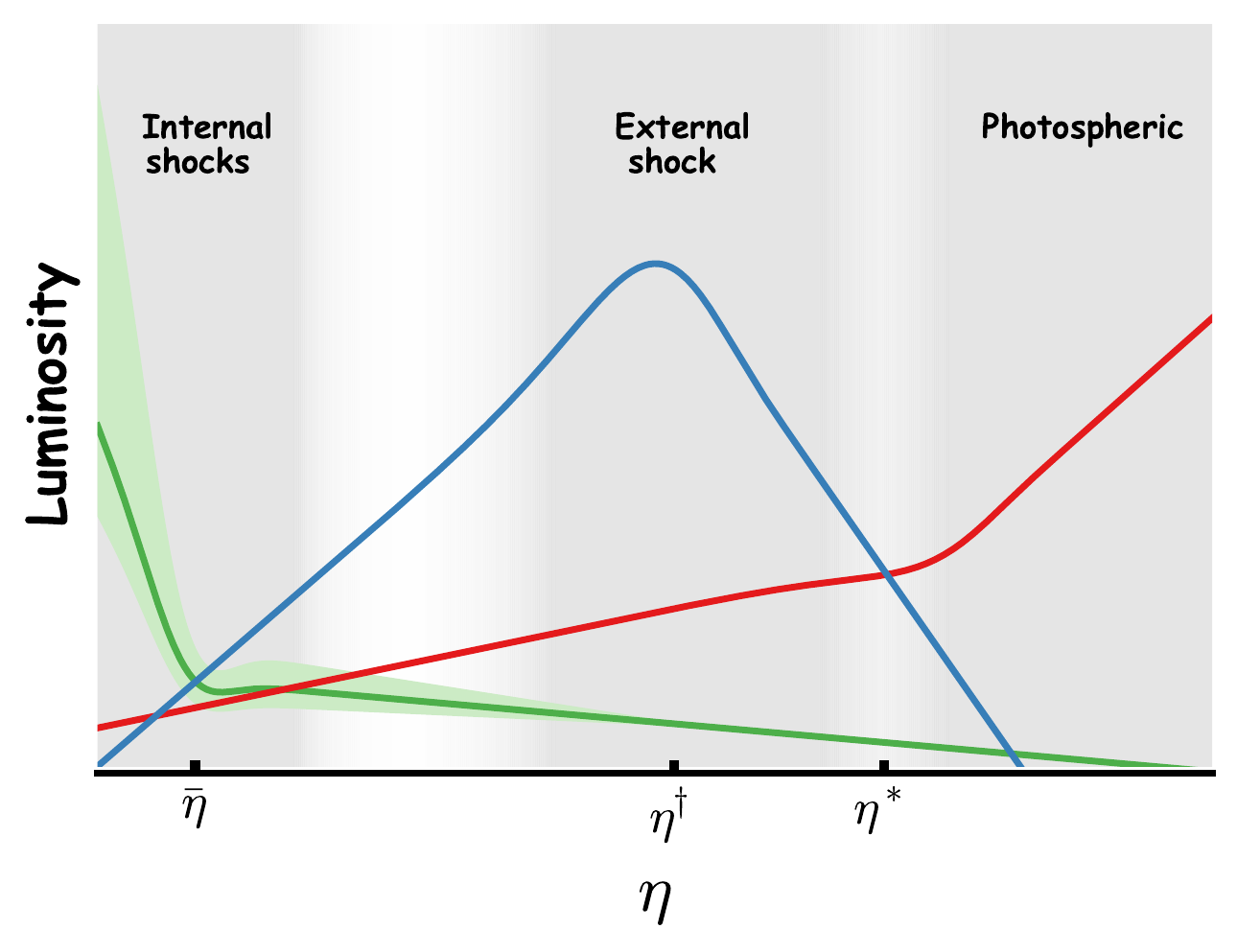}
\caption{Cartoon showing the luminosity of the different emission
  mechanisms in function of the dimensionless entropy. At low $\eta$,
  the emission is dominated by internal shocks (in green). At
  intermediate values of the dimensionless entropy, external shock (in
  blue) becomes the dominant emission mechanism, while at large $\eta$
  the spectrum is expected to be nearly thermal, as mainly originating
  from the photosphere (in red).}
\label{fig:figurecartoon}
\end{figure}

Initially proposed by \cite{RM92}, external shocks were studied in
detail by \cite{CD99}  and \cite{PM98}. As the relativistic blast wave expands, it
substantially slows at the deceleration radius
%& \sim 3.6 \times 10^{16} \left [ \frac{ \left ( \frac{E_{\rm kin, IS}}{10^{52}} \right )}{ n \left ( \frac{\Gamma_{\rm IS}}{100} \right )^2} \right ]^{1/3} \text{cm},

\begin{align}
R_{\rm ES} & = \left (\frac{3 E_{\rm kin, IS}}{4\pi m_p c^2 n \Gamma_{\rm IS} ^2} \right )^{\frac{1}{3}}  \\
& = 1.2\times 10^{17} \text{~~} E_{\rm tot,53}^{\frac{1}{3}} n_0^{-\frac{1}{3}} \epsilon^{-\frac{1}{3}} \eta_{2}^{-\frac{2}{3}} \text{~~}[\text{cm}] \nonumber
\end{align}

% where $E_{\rm kin, IS} = (1-\epsilon_{\rm ph}-\epsilon_{\rm
% IS})E_{\rm tot}=\epsilon E_{\rm tot}$ is the kinetic energy of the
% blast-wave (therefore the remaining kinetic energy after
% photospheric emission and IS),

\noindent where $n$ is the density of the interstellar medium. The
last equality is obtained with $\Gamma_{\rm IS} = \epsilon \eta$,
implied by energy conservation. As the interaction between the outflow
and the CBM develops, two shocks are created: the forward shock which
expands into the CBM, and the reverse shock which collides back into
the outflow. If the reverse shock is not relativistic, the emission
peak time is  \citep{S97}
%\frac{R_{\rm ES}}{2\Gamma_{\rm IS}^2 c} = \frac{1}{2c} \left ( \frac{3}{4 \pi m_p c^2 n} \right )^{\frac{1}{3}} \left [ \frac{E_{\rm tot}}{\eta \Gamma_{\rm IS}^7} \right ]^{\frac{1}{3}} \nonumber \\

\begin{align}
t_{\rm ES} & = \frac{R_{\rm ES}}{\Gamma_{\rm IS}^2 c} =  \left ( \frac{3}{4 \pi m_p c^5 n} \right )^{\frac{1}{3}} \left [ \frac{E_{\rm tot}}{\epsilon^6 \eta^8 } \right ]^{\frac{1}{3}}
\label{eq:delayES}
\end{align} 

% \frac{1}{2c} \left ( \frac{3}{4 \pi m_p c^2 n} \right
% )^{\frac{1}{3}} \left [ \frac{E_{\rm tot}}{\eta \left (
%     1-\frac{\Delta \Gamma}{2 \Gamma_{\rm ph}} \right )^7 \Gamma_{\rm
%   ph}^7} \right ]^{\frac{1}{3}}

\noindent which is strongly dependent on $\eta$. This time-scale also
corresponds to the time delay between  the beginning of the photosphere-IS emissions and the peak of the ES emission\footnote{The delay between IS and PE photons
  is small compared to the long bursts duration and is neglected
  hereafter.}.

%It also scales the luminosity of the ES, the larger $t_{\rm ES}$ the lower the luminosity.

In addition, the luminosity of the forward shock before $t_{\rm ES}$
can be expressed as \citep{S97}

\begin{align}
L_{\rm ES} = \xi_{\rm rad, ES} 32 \pi c^5 n m_p \Gamma_{IS}^8 t^2,
\label{eq:lumES}
\end{align}

\noindent where $t$ is the observed time after trigger and $\xi_{\rm
  rad, ES}$ is the radiative efficiency of the external shock.

%At the deceleration radius, half of the kinetic energy of the blast-wave is extracted. The peak luminosity at $t_{\rm ES}$ can be estimated as follow. Because the area of the shell is increasing, the luminosity can be approximated to grow proportionally to $t^2$ \citep{S97}. Therefore the peak luminosity is given by:
%\begin{align}
%L_{\rm peak} = \frac{3 E_{\rm kin}}{2 t_{\rm ES}}. \label{eq:lumES}
%\end{align}
%As for the IS, the radiative is obtained by multiplying $L_{\rm peak}$ by the radiative efficiency $\xi_{\rm ES}$.

%The weakness of the external shocks is its inability in producing millisecond variability in the light-curve of GRBs. In fact,
%if the CBM density is constant, the light-curve of an ES is made of a single pulse, of typical duration  several seconds.
%Adding variations to the CBM density can produce a light-curve composed of several peaks, however the variability can hardly
%decrease below the second time-scale.

%& \sim \frac{1}{2c} \left ( \frac{E_{\rm tot }}{\eta \left ( 1- \frac{\Delta \Gamma}{2\Gamma_{\rm ph}} \right )} \right )^{\frac{1}{3}} \left ( \frac{1}{\left ( 1- \frac{\Delta \Gamma}{2\Gamma_{\rm ph}} \right )\Gamma_{\rm ph}^2 } \right )^2 \\

%\section{Results: taking into account the time-scale and the efficiency}
\subsection{Thin or thick outflow?}

The typical prompt phase duration of a long GRBs $T_{\rm dur}$ is on
average a few tens of seconds. In the IS and PE framework, this
duration is tightly associated to the light-crossing time of the outflow \citep{DM98,P15}

\begin{align}
T_{\rm dur} = \alpha \frac{l}{c} = \alpha \Delta t.
\end{align}

\noindent Because of redshift dilation, we normalise our computation
to $\Delta t = 5\rm s$\footnote{ Actually, \cite{GGK11} showed that the typical rest-frame duration of \textit{Fermi} bursts is around 12s. Even if normalised to 5s, we find that the following results are weakly dependent on the duration $\Delta t$.}, keeping in mind that it might be much larger.

%where $\alpha$ is a parameter in the order of the unity. 
%Therefore the width of the outflow can be constrained to
%$l \sim 3 \times 10^{11} \alpha (T_{\rm dur}/10\text{s}) \text{cm}$. %This is in line with numerical simulations of jets drilling through
%the progenitor star, for which injection of energy is required on time scale larger than $T_{\rm dur}$.

%In the following, we give a simple estimate of the delay between a photon emitted at the photosphere on the line of sight by the
%inner boundary of the outflow and a photons emitted by the forward shock due to an ES running in the CBM. The origin in time is set for
%a photon emitted at the very beginning of the expansion at the initial radius of expansion and at the time the expansion starts.
%The outer boundary of the outflow is emitted after a time $T_{\rm dur}$. Therefore, the arrival time of the photospheric photon under
%consideration is
%\begin{align}
%t_{\rm a, ph} & = \frac{T_{\rm dur}}{\alpha} + \frac{R_{\rm ph}}{c} - \frac{R_{\rm ph}}{\beta_{\rm ph} c} \nonumber \\
%& =  \frac{T_{\rm dur}}{\alpha} + \frac{R_{\rm ph}}{2\beta_{\rm ph} \Gamma_{\rm ph}^2 c} \text{.}
%\end{align}
%A similar equation can be derived for the IS.

By requiring $t_{\rm ES}$ to be smaller than $\Delta t/2$ such that
photons from the prompt phase originate from all three mechanisms
combined (PE, IS and ES), it follows that:

\begin{align}
  \eta > \eta^{\dagger} & = \left ( \frac{6E_{\rm tot}}{\pi n m_p c^5 \Delta t^3 \epsilon^7} \right )^{\frac{1}{8}} \\
                        & \sim  6.6 \times 10^2 \epsilon^{-\frac{7}{8}} E_{53}^{\frac{1}{8}} n_0^{-\frac{1}{8}} \Delta t_{5\rm s}^{-\frac{3}{8}} \text{}.
\end{align}

Comparing $\eta^{\dagger}$ to $\eta^{*}$ gives a minimum energy
for the burst such that the photosphere does not take place in the
accelerating phase and the external shock peaks in the prompt phase:

\begin{align}
E_{\rm tot} > E_{*} & = 6.9 \times 10^{52} R_{8}^2 \Delta t_{5\rm s}^{2} \epsilon^{-7}  \rm~~ [erg]. \label{eq:minEes}
\end{align}

\noindent The strong dependence on the parameters  has to be noted. Therefore, we
conclude that only very energetic bursts of total energy
$E_{\rm tot} > \text{few~} 10^{53}$ ergs can have a simultaneous
emission from all three mechanisms.

It can be shown that requiring $\eta \gg \eta^{\dagger}$ 
implies that the reverse shock is relativistic \citep{S97}. However, here we use
$t_{\rm ES}$ as a proxy for the peak emission time of the ES. For
instance, density gradients in the ISM below $R_{\rm ES}$ result in
variability of the ES emission before $t_{\rm ES}$, which might also outshine the IS emission.

% REVERSE SHOCK

%\section{Duration, delay and efficiency}
\section{Dominant emission mechanism for a GRB}

\begin{deluxetable*}{c|c|c|c|c}
\tablewidth{0.99\textwidth}
\tablecaption{Spectral and temporal characteristics and/or predictions of GRBs dominated by each three emission mechanisms.\label{tab:predictions}}
\startdata
 &                   &      Internal Shocks        &         External Shocks       &         Photospheric emission \\ \hline 
 &                   &                             &                               &                               \\
Temporal Properties  & Variability        &      Highly variable        &         Smooth pulse          &          Medium to high  variability \\
 &                   &                             &                               &                               \\
                     & Peaks              &      Several                &         One / several very smooth       &         Several  \\ 
 &                   &                             &                               &                               \\
                     & Afterglow          &      Weak and late onset    &         Continuous decay      &         Weak and early onset \\ 
 &                   &                             &                               &                               \\ 
  					 &  Afterglow decay index   &      Can be steep $\delta > 3$    &     $\delta \sim 1.2$      &  Can be steep $\delta > 3$ \\ 
   &                   &                             &                               &                               \\ \hline
Spectral properties  & Thermal component  &       Weak to bright for low $E_{\rm tot}$ &         Weak                  &         Dominant \\ 
 &                   &                             &                               &                               \\
					 & Temporal correlations  &      Yes             &         No                  &         No \\ 
 &                   &                             &                               &                               \\	
 					 & Low energy spectral index  &     $>-2/3$\footnote{This is for synchrotron emission, discarding possible Klein-Nishina effects \citep{DBD11}.}             &        Soft $\alpha \sim -0.67$                  &         Hard $\alpha \sim 0$  \\ 
 &                   &                             &                               &                               \\					 
                     & Pair cut-off       &          $\sim$ 100 MeV          &       $\sim$ 1 GeV            &   -               \\ 
 &                   &                             &                               &                               \\ \hline                     
Other                & Efficiency         &      Low                    &         Middle to high        &         High   
\enddata
\end{deluxetable*}

%%%%%%%%%%%%%%%%%%%%%%%%%%%%%%%%%%%
%%%%%%%%%%%%%%%%%%%%%%%%%%%%%%%%%%%
We now qualitatively analyse the requirements for each emission
mechanism to be the dominant process during the prompt phase of GRBs.

\subsection{Case 1: IS dominates the prompt emission}

% In a situation in which the main component of the prompt emission is
% due to internal shock, let us first remark that the delay between
% the %photospheric emission and an IS scenario is very small
%\begin{align}
%\Delta t_{\rm ph - IS} & \sim \frac{R_{\rm IS} - R_{\rm ph}}{ 2 \Gamma_{\rm ph}^2 c}  \nonumber \\
%& \sim 0.2 \left ( \frac{R_{\rm IS}}{10^{14} \text{cm}} \right ) \left ( \frac{\Gamma_{\rm ph}}{100}  \right )^{-2} \text{s.} 
%\end{align}

%For a long burst, $\Delta t_{\rm ph - IS} \ll l/c$, therefore both
% the photospheric emission and the non-thermal photons are seen
% nearly simultaneously. It is worth noting that the emission from an
% IS scenario is slightly longer than that of the photosphere
% $\alpha_{IS} \gtrsim 1$, as it is not well localised.

In a situation in which the main component of the prompt emission is
due to internal shocks, because of the low efficiency of an internal
shock, three conditions should be met:
\begin{enumerate}
\item the photospheric emission should be dim, that is to say that
  $\Gamma_{\rm ph} \simeq \eta$. From the theory of the photospheric
  emission (see \textit{e.g.} \cite{V14} for a review) it comes that
  lower $\eta$ implies a smaller difference between the Lorentz
  factor at the photosphere and the dimensionless entropy, and also
  a lower brightness of the photosphere.
%\item the outflow should produce an efficient IS. From Equation
%  (\ref{eq:epsilonIS}), it follows that $\Delta \Gamma$ should be as large
%  as possible, compared to $\Gamma_{\rm ph}$.
\item the onset of the ES should take place at late times and it
  should be dim. From Equations (\ref{eq:delayES})
  and (\ref{eq:lumES}), it follows that $\eta$ has to be small.
  %Considering the delay between the
  %duration of the peak of the external shock given by Equation
  %\ref{eq:delayES} and the duration of an internal shock bounded by
  %the light-crossing time of the expanding outflow, it follows that
  %$\Gamma_{\rm ph}$ has to be small.
\end{enumerate}
Therefore, combining the first and last points implies that $\eta$
cannot be too large. If this condition is violated, the IS
emission would be out-shined by the photospheric and/or the ES
emission.
% onset of the emission from the ES would be
% too
% early. %In addition, the most efficient IS are obtained for $\Delta \Gamma$ large enough compared to $\Gamma_{\rm ph} \lesssim \eta$.
To estimate the value of $\eta$ such that the luminosity of the
internal shock is larger than that of the ES at the peak, we combine
Equations (\ref{eq:lumIS}), (\ref{eq:delayES}) and (\ref{eq:lumES}):
% \left( \frac{2}{3^{\frac{2}{3}} c \alpha_{IS} \Delta t } \frac{\xi_{\rm IS}}{\xi_{\rm ES}} \right )^{\frac{3}{8}}  \frac{\epsilon_{\rm IS}^{\frac{3}{8}}}{(1-\epsilon_{\rm IS})^{\frac{5}{4}}} \left ( \frac{E_{\rm tot}}{4\pi m_p c^2 n} \right )^{\frac{1}{8}} \nonumber \\

\begin{align}
\eta < \bar{\eta} & = \frac{1}{\epsilon^{\frac{3}{2}}} \left( \frac{1}{8*3^\frac{2}{3}} \frac{\xi_{\rm rad, IS}}{\xi_{\rm rad, ES}} \frac{\epsilon_{\rm IS}}{\alpha \Delta t} \right )^{\frac{3}{8}} \left ( \frac{E_{\rm tot}}{4\pi m_p c^5 n} \right )^{\frac{1}{8}} \nonumber \\
& = 1.1 \times 10^2 \frac{E_{53}^{\frac{1}{8}}}{\Delta t_{5 \rm s}^{\frac{3}{8}} n_0^{\frac{1}{8}}},
\end{align}

\noindent weakly dependent on the total energy of the burst or on the
CBM density. For the numerical estimate, we choose $\alpha=1$ and $\epsilon =
1-\epsilon_{\rm IS} = 0.8$ (with such a small Lorentz factor, the
photospheric emission is expected to be very dim $\epsilon_{\rm ph}
\ll \epsilon_{\rm IS}, \epsilon_{\rm ES}$). In addition, the radiative
efficiency of the IS and ES are set equal: $\xi_{\rm rad, ES}=\xi_{\rm
  rad, IS}$.

Therefore, an ES should easily outshine the
emission from IS. However, with such a low Lorentz factor
($\bar{\eta}<\eta^{\dagger}$), the ES emission takes place at later
times, leading to a burst composed of two episodes: a precursor followed by one or several smooth
%and a single\footnote{Only if there is no density gradient in the CBM.}
pulses from the external shock at later time.

\subsection{Case 2: ES dominates the prompt emission}
Let us now consider a situation in which the emission from the
external shock dominates the prompt phase, which is only possible if
$E_{\rm tot}>E_{*}$, see Equation (\ref{eq:minEes}). For example, the
prompt phase of GRB~141025A can be interpreted in the ES framework \citep{BBR15}.
The requirements are:
\begin{enumerate}
%\item the IS is dim and not efficient, therefore $\Delta \Gamma \ll
%  \Gamma_{\rm ph}$, which is obtained for large $\Gamma_{\rm ph}$.% or
  %incomplete acceleration, \textit{i.e.} the transparency takes place
  %in the accelerating phase or in the transition phase.
\item the onset of the ES emission should be early. From Equation
  (\ref{eq:delayES}), one sees that it implies large
  $\eta$.% $\Gamma_{\rm ph}$ and thus large $\eta$ (as $\Gamma_{\rm ph} < \eta$).
\item finally, the photospheric emission should not be too bright,
  which implies that the transparency is reached in the coasting
  phase, or in the late transition phase, implying $\Gamma_{\rm ph}
  \lesssim \eta$.
\end{enumerate}
%Therefore, it follows that $\Gamma_{\rm ph}$ should be large, implying
%large $\eta$, to satisfy both the low IS efficiency and the early
%onset of ES. However, $\eta$ can not be \textit{too} large such that the
%transparency takes place in the coasting phase to avoid a too bright
%photosphere: $\eta < \eta^{*}$. %In addition, the burst must be
%energetic enough, with $E_{\rm tot} > E_{*}$ given by Eq.(\ref{eq:minEes}).

 Therefore, it follows that $\eta$ should be large enough to produce the early onset of the afterglow, but small enough such that the transparency takes place in the coasting phase to avoid too bright of a photospheric emission $\eta < \eta^{*}$.
%photosphere: $\eta < \eta^{*}$.} %In addition, the burst must be
%energetic enough, with $E_{\rm tot} > E_{*}$ given by Eq.(\ref{eq:minEes}).

Here, we only considered a circum-burst medium with constant density.
If the density has some variations below $R_{\rm ES}$, a multi-peak
light-curve with \textit{second}-like variability is formed.

\subsection{Case 3: PE the prompt emission}
Finally, if the PE dominates the prompt emission like for GRB~090902B\footnote{Even if this burst is very unique, it perfectly fits in the
  classification scheme.} \citep{RAZ10}, several constraints can also be
obtained
\begin{enumerate}
\item a bright photosphere requires the transparency to take place in
  the accelerating phase or in the transition regime, implying large
  $\eta$ and $\Gamma_{\rm ph} \ll \eta$, that is to say $\eta \sim \eta^{*}$
%\item as in the previous situation, the IS should be dim requiring
%  $\Delta \Gamma \ll \Gamma_{\rm ph}$, which is obtained for large
%  $\Gamma_{\rm ph}$.% or incomplete acceleration (large $\eta$).
\item Finally, a late onset of the ES is required and is obtained for
  low $\Gamma_{\rm ph}$, as implied by the requirement $\eta > \eta^{*}$.% However, the value of $\eta$ does not strongly
  %effect the delay.
\end{enumerate}
%Combining the first and last point implies that $\eta$ should be large
%and $\Gamma_{\rm ph}$ small, which points towards an incomplete
%acceleration at the photosphere.
Therefore, a dominant photosphere requires an incomplete acceleration.

To conclude, we have shown here that for IS to be the dominant
process, low $\eta$ and the transparency in the deep coasting phase
are required. For ES, an intermediate value of $\eta$ with the
transparency reached in the coasting phase are required. Finally, a
spectrum dominated by photospheric emission is obtained if
transparency is reached in the accelerating phase, requiring large
$\eta \gtrsim \eta^{*}$. We note that such results for
PE and ES only were already obtained by
\cite{MRB13}, who considered the photospheric emission as a precursor of the main burst explained by an ES. This is illustrated by a cartoon in Figure
\ref{fig:figurecartoon}.

\section{Additional observational constraints}

The classification scheme proposed here imposes several observational
predictions and requirements, which are summarised by Table
\ref{tab:predictions}. Here we discuss in turn the temporal and
spectral properties of GRBs dominated by one of the three mechanisms.

Both the PE and IS models can
produce highly variable\footnote{The IS model was initially proposed to explain sharp flux decreases on the millisecond time-scale. Such variability is hardly achieved by PE models without fine tuning of the plasma emission at the central engine, and a more realistic variability time-scale might be 0.1-1s.} and complex
(multiples peaks with no correlations between time and the
amplitude/width of a peak) light-curves. However, the ES model cannot easily
explain variability below the second time-scale without fine tuning. In addition,
the ES model can produce several peaks by adding density variations in the
ISM. %But the luminosity of each peak should become shallower
%with time, as the Lorentz factor is decreasing.
However, the duration of each episode should increase and the luminosity
decay should become shallower.
GRB~090618 is an example of such kind of bursts \citep{Z12}: after a first episode
lasting around 50s, which might be associated to PE and/or IS, there are
three episodes with increasing width and decreasing maximum luminosity, which can be
interpreted in the ES framework.

The link between the prompt phase and the ``afterglow" also sets tight
constraints. Indeed a break in luminosity at the very end of the
prompt phase observed by \textit{Swift} cannot easily be explained by an ES,
for which a shallow decay is expected \cite[see however][ who proposed that the decay be explained by a strongly radiative phase triggered by a hadronic discharge]{D07}. % for an
%explanation of the steep decay within the ES framework.
The steep decay might however be characteristic
of an efficient PE or IS model and a low Lorentz factor for the blast
wave after transparency and dissipation by IS, such that the ES
emission is delayed to late times and its luminosity decreased.

The energy requirement to have the ES
occurring simultaneously with PE or IS given by Equation (\ref{eq:minEes}) implies
that a shallow decay of the afterglow, if due to an external shock, is
correlated to large total energy $E_{\rm tot} \gg E_{*}$. This can easily be
checked in the data and is currently under investigation.

Finally, the large difference between $\eta^{*}$ and $\bar{\eta}$
implies that numerous bursts should have (at least) two time-separated components
in their light-curve: first the emission from the photosphere and the IS,
followed by the emission of the ES.

Several constraints can also be obtained from the spectral shape of a burst.
First the relative luminosity of the thermal component and the time evolution of its flux
and temperature, as compared to that of the non-thermal emission can
give clues to identifying the emission mechanism. In particular, in an
IS-PE model, the blackbody properties are likely to track the
non-thermal flux evolution, while it should not be the case in a ES-PE
model.

The rather low Lorentz factor implied if the emission originates from an IS
suggests that a cut-off be present in the spectra at moderate energy of around hundreds
of MeV, see \cite{HDM12}, while larger Lorentz factor as achieved for a dominant
photosphere or ES increases the cut-off to larger energy. The presence or absence
of this cut-off can be investigated with the \textit{Fermi-LAT} instrument.

Finally, the modelling of the afterglow can help to constrain the efficiency
of the prompt emission. Together with its spectral characteristics (specifically
the identification of a blackbody), the emission mechanism might be constrained
as low efficiency is expected from IS, medium from ES and high from PE.

%The idea of two-component GRB light-curves was studied by 

Above, we only mentioned pure cases. However, hybrid bursts are expected to be numerous.
As an example of hybrid, several bursts with envelopes \citep{VMC06} might be explained
by two of the aforementioned emission mechanisms. The univocal determination relies
on a detailed spectral analysis of each component separately.

%BURSTS WITH ENVELOPE PLUS SMALL VARIABILITY

\section{Discussion}

The approach followed in this letter to determine the efficiency of
each emission mechanism is simple, and more detailed computations
can be done. As an example, the ES luminosity and peak energy are
usually determined in the literature by introducing two additional parameters $\epsilon_B$
and $\epsilon_e$, which parametrize the microphysics (magnetic content
and internal energy in random motion of electrons) of the shocked plasma.
Precise values of these parameters are unknown, and limits are often
obtained such that the emission of the forward shock is not detectable
in the MeV band \citep{KB09}, which usually translates to low density $n_0$
and low $\epsilon_B$. However, we note that the limits on the Lorentz
factor do not change substantially with the introduction of 
$\epsilon_B$ and $\epsilon_e$, which justifies our simplified treatment.

Furthermore, our analysis of the photospheric emission is based on
the identification of a black-body. This might be hampered by sub-photospheric
dissipation, \textit{i.e.} below the photosphere, which could result
in strong distortion of the emerging spectrum. Examples of dissipation mechanisms
are neutron decay \citep{B10}, internal shocks \citep{RM05} or dissipation of
magnetic energy \citep{G06}.

Finally, we did not consider outflows where acceleration is powered
by magnetic fields. On one hand, such outflows are challenged by
observations \citep{BP15,BGP15}. On the other hand, it might be possible that
some GRBs outflows be powered by magnetic fields. Theoretically, the dynamics
is parametrized by the magnetisation $\sigma = E_{\rm mag} / Mc^2$, where
$E_{\rm mag}$ is the initial energy in magnetic fields. As for thermally
powered outflows, $\sigma$ plays the same role as $\eta$. In particular, it
scales the coasting Lorentz factor of the outflow. The determination of
spectral and/or temporal criteria to distinguish between thermal and magnetic
acceleration is the next step towards a more detailed classification of
emission mechanisms of GRBs, but it is out of the scope of this paper. 
We also note that if the outflow is strongly magnetized, IS are very inefficient. However, the magnetic field can be dissipated at larger radii by magnetic reconnection, accelerating electrons which radiate synchrotron, see the ICMART model \citep{ZY11}.

To conclude, we have studied the possibility that the three main emission
mechanisms (PE, IS, ES) discussed in the literature be reliable mechanisms
to explain the prompt phase of GRBs. We found that the dimensionless entropy 
and the corresponding Lorentz factors scales the relative luminosity of each mechanisms.
The delay between each emission is also strongly increased with small Lorentz factors,
which implies that bursts with several episodes (in the MeV energy band)
exists. This work presents a simple attempt in classifying the
emission mechanisms of GRBs. 

%\begin{itemize}
%\item $\epsilon_B, \epsilon_e$
%\item subphotospheric dissipation
%\item magnetic outflows
%\item high latitude effects
%\end{itemize}

%\section{Conclusion} 
 
\acknowledgements
The authors are thankful to Peter M\'esz\'aros, Asaf Pe'er, Tsvi Piran and Felix Ryde for discussions which helped improving the manuscript.

DB is supported by a grant from Stiftelsen Olle Engkvist Byggm\"astare.

\bibliographystyle{apj}
%\bibliography{ms.bib}

\begin{thebibliography}{}
\expandafter\ifx\csname natexlab\endcsname\relax\def\natexlab#1{#1}\fi

\bibitem[{{Abbasi} {et~al.}(2012){Abbasi}, {Abdou}, {Abu-Zayyad}, {Ackermann},
  {Adams}, {Aguilar}, {Ahlers}, {Altmann}, {Andeen}, {Auffenberg}, \&
  et~al.}]{AAA12}
{Abbasi}, R., {Abdou}, Y., {Abu-Zayyad}, T., {et~al.} 2012, \nat, 484, 351

\bibitem[{{Ahlgren} {et~al.}(2015), {Ahlgren}, {Larsson}, {Nymark}, {Ryde}, \& {Pe'er}}]{ALN15}
{Ahlgren}, B. and {Larsson}, J. and {Nymark}, T. and {Ryde}, F. and {Pe'er}, A. 2015, \mnras, 454, 31

\bibitem[{{B{\'e}gu{\'e}} \& {Pe'er}(2015)}]{BP15}
{B{\'e}gu{\'e}}, D., \& {Pe'er}, A. 2015, \apj, 802, 134

\bibitem[{{B{\'e}gu{\'e}} \& {Iyyani}(2014)}]{BI14}
{B{\'e}gu{\'e}}, D., \& {Iyyani}, S. 2014, \apj, 792, 42


\bibitem[{{Beloborodov}(2010)}]{B10}
{Beloborodov}, A.~M. 2010, \mnras, 407, 1033

\bibitem[{{Bromberg} {et~al.}(2015){Bromberg}, {Granot}, \& {Piran}}]{BGP15}
{Bromberg}, O., {Granot}, J., \& {Piran}, T. 2015, \mnras, 450, 1077

\bibitem[{{Burgess} {et~al.}(2015){Burgess}, {B{\'e}gu{\'e}}, {Ryde}, {Omodei},
  {Pe'er}, {Racusin}, \& {Cucchiara}}]{BBR15}
{Burgess}, J.~M., {B{\'e}gu{\'e}}, D., {Ryde}, F., {et~al.} 2015, ArXiv
  e-prints, arXiv:1506.05131

\bibitem[{{Chiang} \& {Dermer}(1999)}]{CD99}
{Chiang}, J., \& {Dermer}, C.~D. 1999, \apj, 512, 699

\bibitem[{{Daigne} {et~al.}(2011){Daigne}, {Bo\v{s}njak}, \& {Dubus}}]{DBD11}
{Daigne}, F., {Bo\v{s}njak}, v., \& {Dubus}, G. 2011, \aap, 526, A110

\bibitem[{{Daigne} \& {Mochkovitch}(1998)}]{DM98}
{Daigne}, F., \& {Mochkovitch}, R. 1998, \mnras, 296, 275

\bibitem[{{Daigne} \& {Mochkovitch}(2002)}]{DM02}
---. 2002, \mnras, 336, 1271

\bibitem[{{Dermer}(2007)}]{D07}
{Dermer}, C.~D. 2007, \apj, 664, 384

\bibitem[{{Drenkhahn} \& {Spruit}(2002)}]{DS02}
{Drenkhahn}, G., \& {Spruit}, H.~C. 2002, \aap, 391, 1141

\bibitem[{{Ghirlanda} {et al.}(2003) {Ghirlanda}, {Celotti} \&  {Ghisellini}}]{GCG03}
{Ghirlanda}, G. and {Celotti}, A. and {Ghisellini}, G. 2003, \aap, 406, 879


\bibitem[{{Giannios}(2006)}]{G06}
{Giannios}, D. 2006, \aap, 457, 763

\bibitem[{{Golkhou} \& {Butler}(2014)}]{GB14}
{Golkhou}, V.~Z., \& {Butler}, N.~R. 2014, \apj, 787, 90

\bibitem[{{Goodman}(1986)}]{G86}
{Goodman}, J. 1986, \apjl, 308, L47

\bibitem[{{Grimsrud} \& {Wasserman}(1998)}]{GW98}
{Grimsrud}, O.~M., \& {Wasserman}, I. 1998, \mnras, 300, 1158

\bibitem[{{Gruber} {et~al.}(2011){Gruber}, {Greiner}, {von Kienlin}, {Rau},
	{Briggs}, {Connaughton}, {Goldstein}, 
	{van der Horst}, {Nardini}, {Bhat}, 
	{Bissaldi}, {Burgess}, {Chaplin}, 
	{Diehl}, {Fishman}, {Fitzpatrick}, {Foley}, 
	{Gibby}, {Giles}, {Guiriec}, {Kippen}, 
	{Kouveliotou}, {Lin}, {McBreen}, {Meegan}, 
	{Olivares E.}, {Paciesas}, {Preece}, 
	{Tierney}, {Wilson-Hodge}}]{GGK11}
{Gruber}, D., {Greiner}, J., {von Kienlin}, A., {et~al.} 2011, \aap, 531, A20


\bibitem[{{Hasco{\"e}t} {et~al.}(2012){Hasco{\"e}t}, {Daigne}, {Mochkovitch},
  \& {Vennin}}]{HDM12}
{Hasco{\"e}t}, R., {Daigne}, F., {Mochkovitch}, R., \& {Vennin}, V. 2012,
  \mnras, 421, 525

\bibitem[{{Iyyani} {et~al.}(2013){Iyyani}, {Ryde}, {Axelsson}, {Burgess}, 
	{Guiriec}, {Larsson}, {Lundman}, {Moretti}, 
	{McGlynn}, {Nymark}, \& {Rosquist}}]{IRA13}
{Iyyani}, S., {Ryde}, F., {Axelsson}, M. , {et~al.} 2013, \mnras, 433, 2739

\bibitem[{{Kobayashi} {et~al.}(1997){Kobayashi}, {Piran}, \& {Sari}}]{KPS97}
{Kobayashi}, S., {Piran}, T., \& {Sari}, R. 1997, \apj, 490, 92

\bibitem[{{Kumar} \& {Barniol Duran}(2009)}]{KB09}
{Kumar}, P., \& {Barniol Duran}, R. 2009, \mnras, 400, L75

\bibitem[{{Kumar} \& {Zhang}(2015)}]{KZ15}
{Kumar}, P., \& {Zhang}, B. 2015, \physrep, 561, 1

\bibitem[{{Lazzati} {et al.}(2012), {Morsony}, {Blackwell}, \& {Begelman}}]{LMB12}
{Lazzati}, D. and {Morsony}, B.~J. and {Blackwell}, C.~H. and {Begelman}, M.~C. 2012, \apj, 750, 68

\bibitem[{{M{\'e}sz{\'a}ros} (2006)}]{M06}
{M{\'e}sz{\'a}ros}, P. 2006, Reports on Progress in Physics, 69, 2259

\bibitem[{{M{\'e}sz{\'a}ros} \& {Rees}(1993)}]{RM93}
{M{\'e}sz{\'a}ros}, P., \& {Rees}, M.~J. 1993, \apj, 405, 278

\bibitem[{{M{\'e}sz{\'a}ros} \& {Rees}(2000)}]{MR00}
---. 2000, \apj, 530, 292

\bibitem[{{Muccino} {et~al.}(2013){Muccino}, {Ruffini}, {Bianco}, {Izzo}, \& {Penacchioni}}]{MRB13}
{Muccino}, M., {Ruffini}, R., {Bianco}, C.~L., {Izzo}, L., \& {Penacchioni}, A.~V. 2013, \apj, 763, 125

\bibitem[{{Narayan} {et~al.}(2007){Narayan}, {McKinney}, \& {Farmer}}]{NMF07}
{Narayan}, R., {McKinney}, J.~C., \& {Farmer}, A.~J. 2007, \mnras, 375, 548

\bibitem[{{Paczy{\'n}ski}(1986)}]{P86}
{Paczy{\'n}ski}, B. 1986, \apjl, 308, L43

\bibitem[{{Paczy{\'n}ski}(1990)}]{P90}
---. 1990, \apj, 363, 218

\bibitem[{{Panaitescu} \& {M{\'e}sz{\'a}ros}(1998)}]{PM98}
{Panaitescu}, A., \& {M{\'e}sz{\'a}ros}, P. 1998, \apj, 492, 683

\bibitem[{{Pe'er}(2008)}]{P08}
{Pe'er}, A. 2008, \apj, 682, 463

\bibitem[{{Pe'er}(2015)}]{P15}
{Pe'er}, A. 2015, Advances in Astronomy, 2015, 907321


\bibitem[{{Piran} \& {Shemi}(1993)}]{PS93}
{Piran}, T., \& {Shemi}, A. 1993, \apjl, 403, L67

\bibitem[{{Piran} {et~al.}(1993){Piran}, {Shemi}, \& {Narayan}}]{PSN93}
{Piran}, T., {Shemi}, A., \& {Narayan}, R. 1993, \mnras, 263, 861

\bibitem[{{Rees} \& {M{\'e}sz{\'a}ros}(1992)}]{RM92}
{Rees}, M.~J., \& {M{\'e}sz{\'a}ros}, P. 1992, \mnras, 258, 41P

\bibitem[{{Rees} \& {M{\'e}sz{\'a}ros}(1994)}]{RM94}
---. 1994, \apjl, 430, L93

\bibitem[{{Rees} \& {M{\'e}sz{\'a}ros}(2005)}]{RM05}
---. 2005, \apj, 628, 847

\bibitem[{{Ryde}(2004)}]{R04}
{Ryde}, F. 2004, \apj, 614, 827

\bibitem[{{Ryde} {et~al.}(2010){Ryde}, {Axelsson}, {Zhang}, {McGlynn}, {Pe'er},
  {Lundman}, {Larsson}, {Battelino}, {Zhang}, {Bissaldi}, {Bregeon}, {Briggs},
  {Chiang}, {de Palma}, {Guiriec}, {Larsson}, {Longo}, {McBreen}, {Omodei},
  {Petrosian}, {Preece}, \& {van der Horst}}]{RAZ10}
{Ryde}, F., {Axelsson}, M., {Zhang}, B.~B., {et~al.} 2010, \apjl, 709, L172

\bibitem[{{Rykoff} {et~al.}(2009){Rykoff}, {Aharonian}, {Akerlof}, {Ashley},
  {Barthelmy}, {Flewelling}, {Gehrels}, {G{\"o}\v{g}{\"u}\c{s}}, {G{\"u}ver},
  {Kizilo\v{g}lu}, {Krimm}, {McKay}, {{\"O}zel}, {Phillips}, {Quimby},
  {Rowell}, {Rujopakarn}, {Schaefer}, {Smith}, {Vestrand}, {Wheeler}, {Wren},
  {Yuan}, \& {Yost}}]{RAA09}
{Rykoff}, E.~S., {Aharonian}, F., {Akerlof}, C.~W., {et~al.} 2009, \apj, 702,
  489

\bibitem[{{Sari}(1997)}]{S97}
{Sari}, R. 1997, \apjl, 489, L37

\bibitem[{{Thompson}(1994)}]{T94}
{Thompson}, C. 1994, \mnras, 270,480

\bibitem[{{Thompson} {et al.}(1994){Thompson}, {M{\'e}sz{\'a}ros} \& {Rees}}]{TMR07}
{Thompson}, C. and {M{\'e}sz{\'a}ros}, P. and {Rees}, M.~J. 2007, \apj, 666, 1012


\bibitem[{{Vereshchagin}(2014)}]{V14}
{Vereshchagin}, G.~V. 2014, International Journal of Modern Physics D, 23,
  30003

\bibitem[{{Vetere} {et~al.}(2006){Vetere}, {Massaro}, {Costa}, {Soffitta}, \&
  {Ventura}}]{VMC06}
{Vetere}, L., {Massaro}, E., {Costa}, E., {Soffitta}, P., \& {Ventura}, G.
  2006, \aap, 447, 499

\bibitem[{{Woosley}(1993)}]{W93}
{Woosley}, S.~E. 1993, \apj, 405, 273

\bibitem[{{Woosley}\& {Bloom}(2006)}]{WB06}
{Woosley}, S.~E., \& {Bloom}, J.~S. 2006, \araa, 44, 507


\bibitem[{{Zhang} \& {Yan}(2011)}]{ZY11}
{Zhang}, B., \& {Yan}, H. 2011, \apj, 726, 90

\bibitem[{{Zhang}(2012)}]{Z12}
{Zhang}, F.-W. 2012, \apss, 339, 123

\end{thebibliography}

\end{document}